\pdfoutput=1
\documentclass[preprint,preprintnumbers,amsmath,amssymb,floatfix,endfloats*]{revtex4}

\usepackage{graphicx}
\usepackage{dcolumn}
\usepackage{bm}
\usepackage{amsfonts}

\DeclareMathAlphabet{\mathsfsl}{OT1}{cmr}{bx}{it}
\begin{document}
\title{Accelerated relaxation in disordered solids under cyclic loading with alternating shear orientation}
\author{Nikolai V. Priezjev$^{1,2}$}
\affiliation{$^{1}$Department of Mechanical and Materials
Engineering, Wright State University, Dayton, OH 45435}
\affiliation{$^{2}$National Research University Higher School of
Economics, Moscow 101000, Russia}
\date{\today}
\begin{abstract}

The effect of alternating shear orientation during cyclic loading on
the relaxation dynamics in disordered solids is examined using
molecular dynamics simulations.  The model glass was initially
prepared by rapid cooling from the liquid state and then subjected
to cyclic shear along a single plane or periodically alternated in
two or three dimensions.  We showed that with increasing strain
amplitude in the elastic range, the system is relocated to deeper
energy minima. Remarkably, it was found that each additional
alternation of the shear orientation in the deformation protocol
brings the glass to lower energy states. The results of mechanical
tests after more than a thousand shear cycles indicate that cyclic
loading leads to the increase in strength and shear-modulus
anisotropy.

\vskip 0.5in


\end{abstract}

\maketitle

\section{Introduction}

Understanding the response of metallic glasses to mechanical and
thermal loading is important for a broad range of applications in
biomedical engineering and environmental science~\cite{Ruta17}.
Unlike crystalline materials, where plastic deformation can be
described in terms of topological line defects, the amorphous
structure of glasses lacks long-range order and can be changed via
collective rearrangements of small groups of
atoms~\cite{Spaepen77,Argon79}. During thermomechanical processing,
the glasses can be brought to either higher energy, rejuvenated
states or to lower energy, relaxed states~\cite{Greer16}. The
rejuvenated samples are typically less brittle, whereas relaxed
glasses exhibit high yield stress and greater chemical stability,
among other advantages~\cite{Greer16}. There are a number of
experimental techniques that can be used to increase the stored
energy; namely, high pressure torsion, ion irradiation, wire
drawing, and shot peening. Moreover, it was shown that prolonged
static loading below the yield stress can induce either irreversible
structural rearrangements or the formation of nanocrystals in the
metallic glass matrix, leading to improved
plasticity~\cite{Park08,Eckert11,Tong13,Gu14,GreerSun16,PanGreer18,PriezELAST19}.
Alternatively, aged glasses can be rejuvenated by recovery annealing
above the glass transition temperature and then cooled down with a
suitably fast rate~\cite{Saida13,Ogata15,Maass18,Yang18,Priez19one}.
The mechanical and magnetic properties can also be tuned by repeated
thermal cycling at ambient pressure, which might result in
rejuvenation or relaxation depending on the system size and thermal
amplitude~\cite{Ketov15,Bu16,Ri17,LiuNano18,Priez19tcyc,Priez19T2000,Priez19T5000,Kruzic19}.
Nevertheless, the combined effect of periodic mechanical deformation
and thermal cycling on the extent of relaxation and rejuvenation
remains largely unexplored.

\vskip 0.05in

In recent years, a number of groups have studied the structural
relaxation and dynamical properties of disordered solids under
periodic loading using atomistic simulations~\cite{Priezjev13,
Sastry13,Reichhardt13,Priezjev14,IdoNature15,Priezjev16,Kawasaki16,
Priezjev16a,Sastry17,Priezjev17,OHern17,Hecke17,Priezjev18,
Priezjev18a,NVP18strload,Sastry18,PriMakrho05,PriMakrho09}.  In
particular, it was found that during cyclic shear along a single
plane, amorphous systems relocate to lower energy states via a
sequence of collective, irreversible rearrangements of atoms when
the strain amplitude is below the yielding strain~\cite{Priezjev13,
Sastry13,Reichhardt13,Priezjev14,IdoNature15,Priezjev16,Sastry17,
Priezjev18,Priezjev18a,Sastry18,PriMakrho05,PriMakrho09}.  In the
case of athermal quasistatic loading, the disordered systems
eventually reach a special dynamical state, the so-called limit
cycle, where the particle dynamics becomes exactly reversible after
one or several
cycles~\cite{Sastry13,Reichhardt13,IdoNature15,Sastry17}. At larger
strain amplitudes, the yielding transition typically occurs after a
number of transient cycles and it is accompanied by the formation of
a shear band where the flow is localized and the potential energy is
enhanced~\cite{Sastry17,Priezjev17,Priezjev18a}. More recently, it
was shown that shearing along a single or alternating planes did not
result in significant difference of the potential energy when the
strain amplitude is close to the critical value~\cite{Sastry18}.
However, despite extensive efforts to access low energy states via
mechanical agitation, the development of more efficient deformation
protocols remains a challenge.

\vskip 0.05in

In this paper, the influence of cyclic deformation with alternating
orientation of the shear plane on structural relaxation in binary
glasses is investigated via molecular dynamics simulations. We
consider a poorly annealed glass at a temperature well below the
glass transition temperature and impose periodic shear strain
deformation for over a thousand cycles.  It will be shown that, in
general, glasses relocate to lower energy states with increasing
strain amplitude in the elastic range. Moreover, for a given strain
amplitude, the relaxation is accelerated when an additional shear
orientation is implemented in the deformation protocol.  As a result
of cyclic loading, the relaxed glasses exhibit an increase in
strength and elastic stress anisotropy.

\vskip 0.05in

The rest of the paper is divided into three sections. The
description of the model, preparation procedure, and deformation
protocols is given in the next section.  The analysis of the
potential energy series and mechanical properties is presented in
section\,\ref{sec:Results}.  The last section contains a brief
summary.

\section{Molecular dynamics (MD) simulations}
\label{sec:MD_Model}

The model glass is represented via the binary (80:20) mixture with
strongly non-additive interaction between different types of atoms,
which prevents crystallization~\cite{KobAnd95}. The Kob-Andersen
(KA) model was extensively studied in the last twenty years and it
is often used as a benchmark for testing novel algorithms to examine
statistical properties of glassy systems~\cite{KobAnd95}. In this
model, the interaction between atoms $\alpha,\beta=A,B$ is described
via the Lennard-Jones (LJ) potential:
\begin{equation}
V_{\alpha\beta}(r)=4\,\varepsilon_{\alpha\beta}\,\Big[\Big(\frac{\sigma_{\alpha\beta}}{r}\Big)^{12}\!-
\Big(\frac{\sigma_{\alpha\beta}}{r}\Big)^{6}\,\Big],
\label{Eq:LJ_KA}
\end{equation}
with the following parametrization: $\varepsilon_{AA}=1.0$,
$\varepsilon_{AB}=1.5$, $\varepsilon_{BB}=0.5$, $\sigma_{AA}=1.0$,
$\sigma_{AB}=0.8$, $\sigma_{BB}=0.88$, and
$m_{A}=m_{B}$~\cite{KobAnd95}.  We simulate a relatively large
system of $60\,000$ atoms and use the cutoff radius
$r_{c,\,\alpha\beta}=2.5\,\sigma_{\alpha\beta}$ to reduce the
computational efforts.  In the following, unless otherwise
specified, the MD results are reported in the LJ units of length,
mass, energy, and time: $\sigma=\sigma_{AA}$, $m=m_{A}$,
$\varepsilon=\varepsilon_{AA}$, and
$\tau=\sigma\sqrt{m/\varepsilon}$.  The simulations were carried out
using the LAMMPS parallel code~\cite{Lammps} with the time step
$\triangle t_{MD}=0.005\,\tau$.

\vskip 0.05in


The binary mixture was first thoroughly equilibrated at the
temperature $T_{LJ}=1.0\,\varepsilon/k_B$ and density
$\rho=\rho_A+\rho_B=1.2\,\sigma^{-3}$. Here, the parameters $k_B$
and $T_{LJ}$ stand for the Boltzmann constant and temperature,
respectively. The system temperature was controlled via the
Nos\'{e}-Hoover thermostat~\cite{Allen87,Lammps}. Note that periodic
boundary conditions were imposed along all three spatial dimensions.
The corresponding box size at the density $\rho=1.2\,\sigma^{-3}$ is
$L=36.84\,\sigma$. The critical temperature of the KA model at this
density is $T_c=0.435\,\varepsilon/k_B$~\cite{KobAnd95}. After
equilibration, the system was linearly cooled to the very low
temperature $T_{LJ}=0.01\,\varepsilon/k_B$ with the fast rate of
$10^{-2}\varepsilon/k_{B}\tau$ at constant volume.

\vskip 0.05in


Next, the glass was subjected to periodic shear strain deformation
with the oscillation period $T=5000\,\tau$ and frequency
$\omega=2\pi/T=1.26\times10^{-3}\,\tau^{-1}$, as follows:
\begin{equation}
\gamma(t)=\gamma_0\,\text{sin}(2\pi t/T),
\label{Eq:shear}
\end{equation}
where $\gamma_0$ is the strain amplitude. The deformation protocol
consists of either periodic shear along the $xz$ plane, or
alternating shear along the $xz$ and $yz$ planes, or alternating
shear along the $xz$, $yz$, and $xy$ planes. In all cases, the
sample was deformed during 1400 shear cycles with strain amplitudes,
$\gamma_0\leqslant0.065$, in the elastic range at
$T_{LJ}=0.01\,\varepsilon/k_B$ and $\rho=1.2\,\sigma^{-3}$. After
cyclic loading, the samples were relaxed during $10^4\,\tau$ at zero
strain and $T_{LJ}=0.01\,\varepsilon/k_B$, and then sheared with the
rate $10^{-5}\,\tau^{-1}$ at constant volume along the $xz$, $yz$,
and $xy$ planes.  The shear modulus and the maximum value of the
shear stress were computed from the stress-strain response at
constant strain rate. In addition, the potential energy, system
dimensions, stress components, and atomic configurations were saved
for the post-processing analysis.  A typical MD simulation of 1400
shear cycles required about 900 hours using 40 processors. Due to
computational limitations, the simulations of periodic shear
deformation were performed only for one sample.

\section{Results}
\label{sec:Results}


It is well known that in amorphous materials at very low
temperatures, the motion of atoms mostly involves vibration within
cages formed by their neighbors, and, as a result, the structural
relaxation is strongly suppressed. One way to induce collective
rearrangements of atoms and to access lower energy states is to
periodically deform the material with the strain amplitude below the
yielding point~\cite{Lacks04}.  Under applied strain, the potential
energy landscape becomes tilted, and a group of atoms might relocate
to a lower energy minimum. A number of recent MD studies have shown
that under cyclic shear along a single plane, disordered solids are
driven to lower energy states via irreversible rearrangements of
atoms~\cite{Priezjev13,Sastry13,Reichhardt13,Priezjev16,Sastry17,Priezjev18,Priezjev18a}.
Moreover, it was found that athermal systems eventually reach a
fully reversible limit cycle and settle down at a particular energy
level~\cite{Sastry13,Reichhardt13,IdoNature15,Sastry17}, while in
the presence of thermal fluctuations, glasses continue to explore
lower energy states, although the size of clusters of mobile atoms
becomes progressively
smaller~\cite{Priezjev18,Priezjev18a,NVP18strload}.


\vskip 0.05in

The slowing down of the relaxation dynamics during cyclic shear
along a single plane implies the existence of the lowest bound in
energy that can be attained using such deformation protocol. On the
other hand, if the direction of the cyclic shear plane is changed
after a number of cycles, then the shear strain in the new direction
might allow for larger-scale rearrangements and, consequently,
relocation to deeper energy states.  It remains an open question
whether a fully reversible state with a lower potential energy can
be reached in athermal systems using a combination of shear cycles
oriented along different directions.   In the present study, we
compare three deformation protocols where the cyclic shear is either
oriented along a single plane or periodically alternated in two and
three dimensions. The simulations are performed at strain amplitudes
below the yielding transition.  It was recently shown using athermal
simulations that the critical strain amplitude of the KA mixture at
the density $\rho=1.2\,\sigma^{-3}$ is in the range $0.07< \gamma_0
< 0.08$, and it might depend on the system size, since the
nucleation of a shear band is determined by its total interfacial
area~\cite{Sastry13,Sastry17}.   In the presence of thermal
fluctuations, the critical strain amplitude is reduced with
increasing system
temperature~\cite{Priezjev13,Priezjev18,Priezjev18a}.

\vskip 0.05in

\begin{figure}[t]
\includegraphics[width=12.0cm,angle=0]{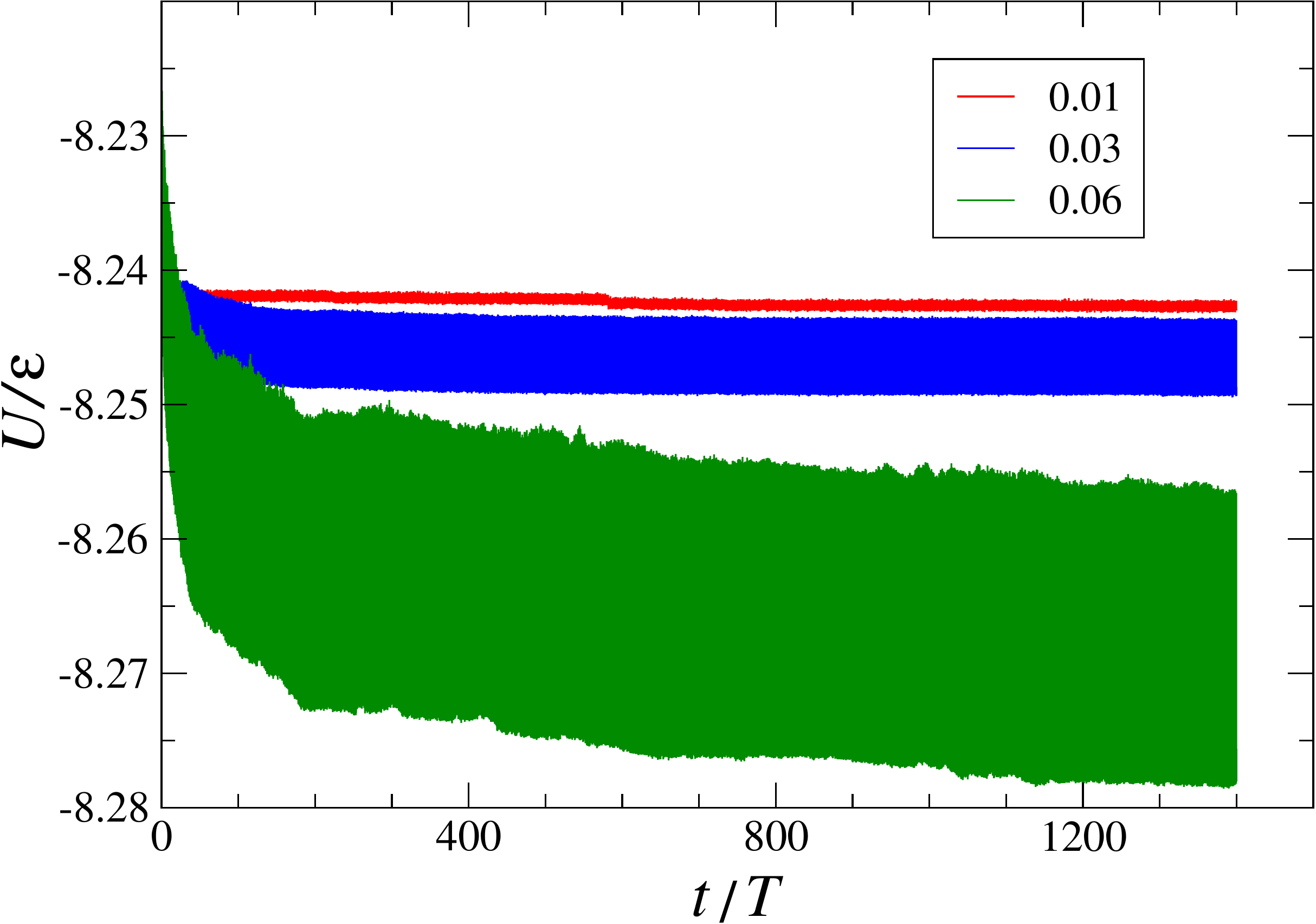}
\caption{(Color online) The potential energy series during 1400
shear cycles along the $xz$ plane. The strain amplitudes are
$\gamma_0=0.01$, $0.03$, and $0.06$ (from top to bottom). The period
of oscillation is $T=5000\,\tau$.}
\label{fig:poten_amp01_03_06}
\end{figure}

The dependence of the potential energy as a function of time during
periodic shear along a single plane ($xz$) is shown in
Fig.\,\ref{fig:poten_amp01_03_06} for the strain amplitudes
$\gamma_0=0.01$, $0.03$, and $0.06$. Here, the time is expressed in
oscillation periods.   It can be clearly observed that with
increasing strain amplitude, the system relocates to deeper energy
minima and the amplitude of potential energy oscillations becomes
larger. At lower strain amplitudes, $\gamma_0=0.01$ and $0.03$, the
potential energy gradually approaches nearly constant levels via
occasional sudden drops, which are associated with irreversible
rearrangements of groups of atoms. Notice, for example, the step at
$t\approx580\,T$ for $\gamma_0=0.01$ in
Fig.\,\ref{fig:poten_amp01_03_06}.  By contrast, the potential
energy at $\gamma_0=0.06$ is significantly lower and it is subject
to large fluctuations, which occur due to the finite temperature and
$\gamma_0$ being sufficiently close to the critical value. Thus, it
was shown that the potential energy minima approach a constant level
after about 100 shear cycles at $\gamma_0=0.06$, when a poorly
annealed glass was periodically deformed using the athermal
quasistatic protocol~\cite{Sastry17}.

\vskip 0.05in

\begin{figure}[t]
\includegraphics[width=12.0cm,angle=0]{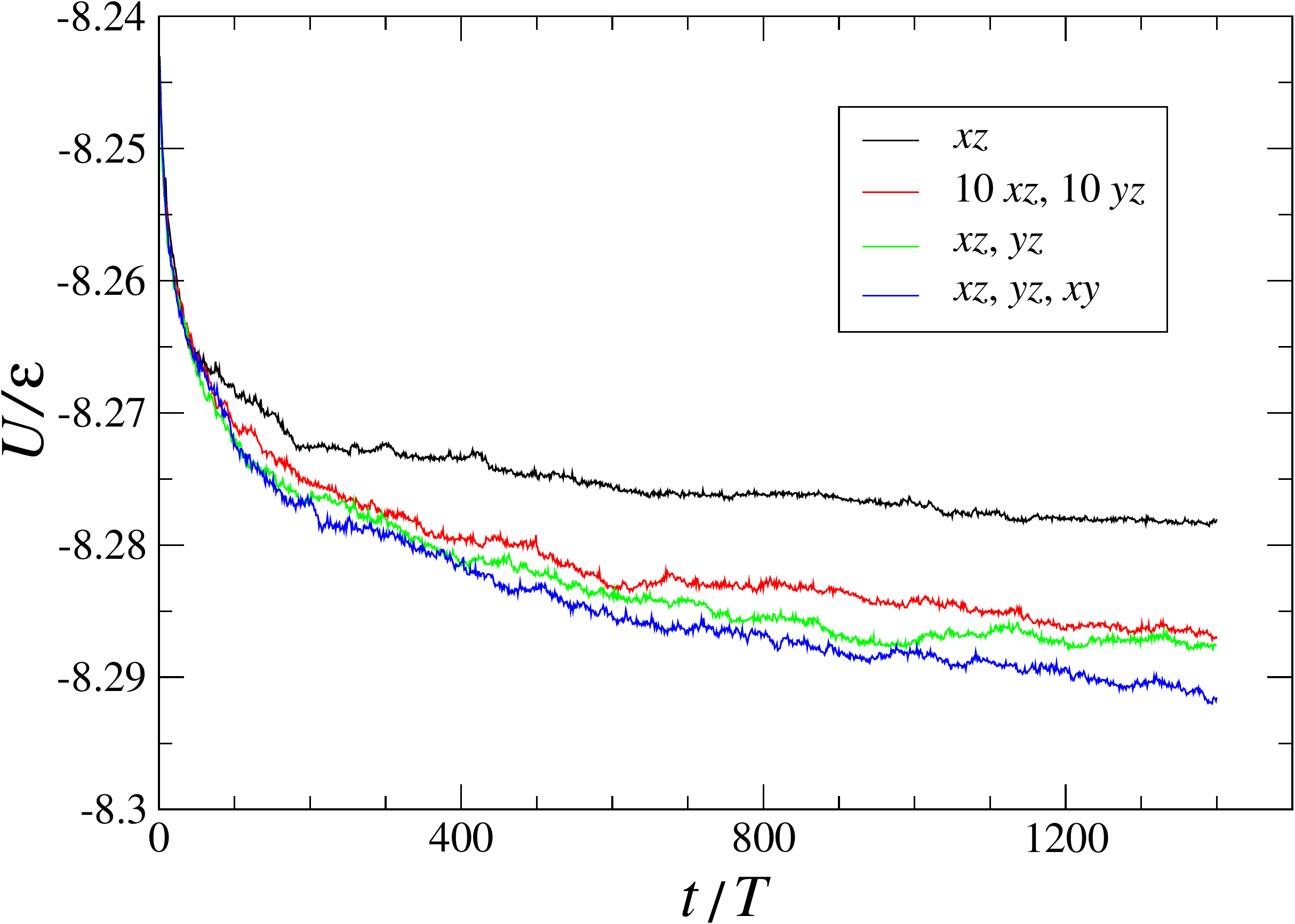}
\caption{(Color online) The potential energy minima during 1400
shear cycles for the strain amplitude $\gamma_0=0.06$. The
deformation protocols include: (1) periodic shear along the $xz$
plane, (2) 10 shear cycles along the $xz$ plane, followed by 10
shear cycles along the $yz$ plane, (3) alternating shear along the
$xz$ and $yz$ planes, and (4) alternating shear along the $xz$,
$yz$, and $xy$ planes, as indicated in the legend. The oscillation
period is $T=5000\,\tau$ and the temperature is
$T_{LJ}=0.01\,\varepsilon/k_B$.  }
\label{fig:poten_amp06_4prs}
\end{figure}

Next, the effect of alternating shear orientation on the potential
energy minima is demonstrated in Fig.\,\ref{fig:poten_amp06_4prs}
for different deformation protocols at the strain amplitude
$\gamma_0=0.06$.   In the case of periodic shear along the $xz$
plane at $\gamma_0=0.06$, shown in
Fig.\,\ref{fig:poten_amp01_03_06}, only the potential energies at
the end of each cycle are replotted in
Fig.\,\ref{fig:poten_amp06_4prs}.   The other three deformation
protocols involve (a) 10 consecutive shear cycles along the $xz$
plane, followed by 10 shear cycles along the $yz$ plane, (b)
alternating shear along the $xz$ and $yz$ planes, and (c)
alternating shear along the $xz$, $yz$, and $xy$ planes. The results
in Fig.\,\ref{fig:poten_amp06_4prs} indicate that each additional
alternation of the shear orientation in the deformation protocol
leads to lower energy states. In particular, it can be seen that the
potential energy minima become significantly lower even when the
shear is applied along the same plane for a number of cycles before
its orientation is changed (every 10 cycles). Among all protocols,
the lowest energy ($U\approx-8.29\,\varepsilon$) is attained in the
limiting case when the orientation of the shear plane is alternated
after every cycle in all three spatial dimensions. For comparison,
when the KA mixture is cooled with the slow rate
$10^{-5}\varepsilon/k_{B}\tau$ at $\rho=1.2\,\sigma^{-3}$, the
energy is $U\approx-8.31\,\varepsilon$~\cite{Priezjev17}.

\vskip 0.05in

\begin{figure}[t]
\includegraphics[width=12.0cm,angle=0]{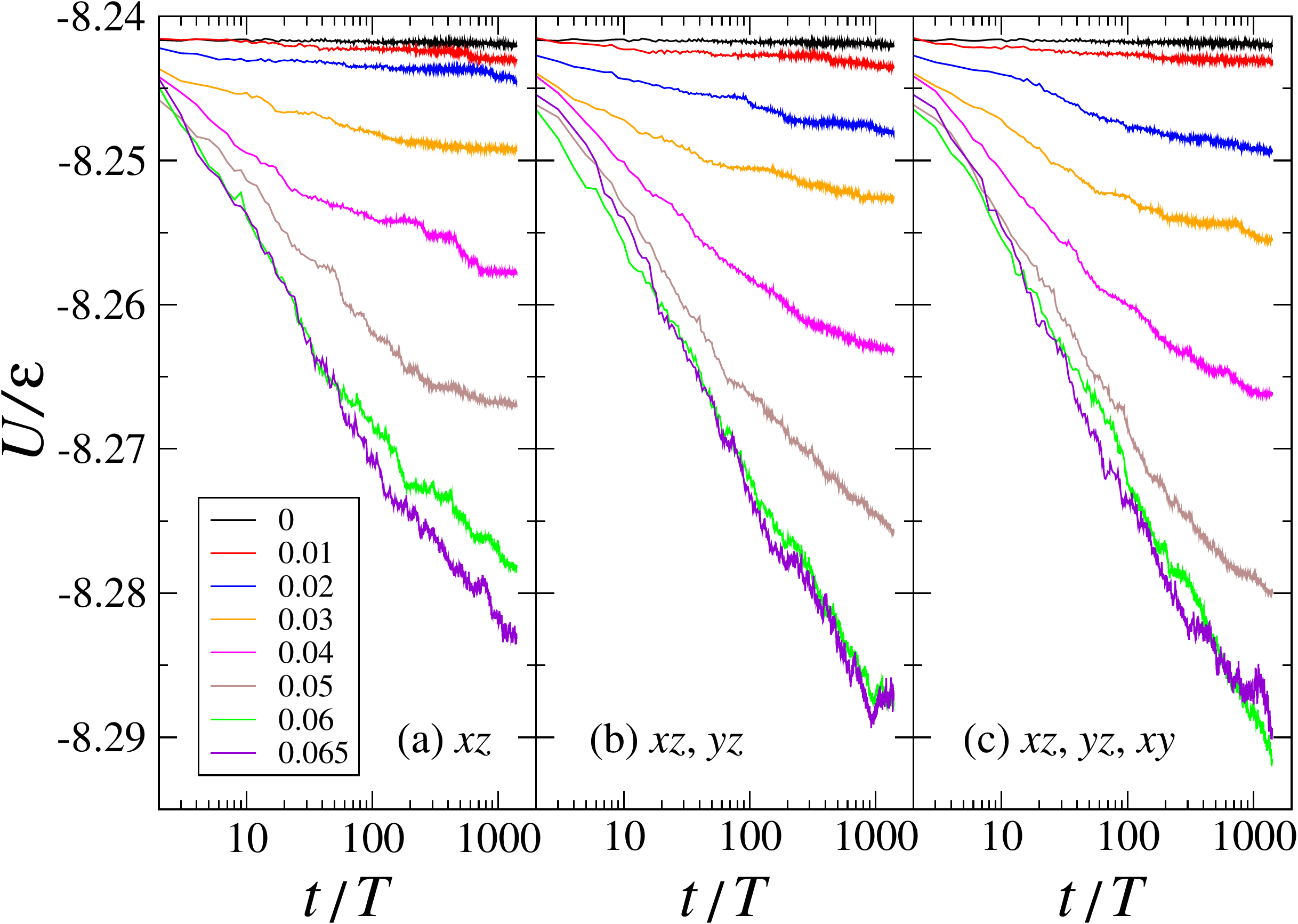}
\caption{(Color online) The potential energy at the end of each
shear cycle for the indicated strain amplitudes. The deformation is
(a) periodic shear along the $xz$ plane, (b) alternating shear along
the $xz$ and $yz$ planes, and (c) alternating shear along the $xz$,
$yz$, and $xy$ planes. The period is $T=5000\,\tau$.  }
\label{fig:poten_amp_3pan}
\end{figure}

The summary of the potential energy during 1400 cycles for three
different protocols is presented in Fig.\,\ref{fig:poten_amp_3pan}
for the strain amplitudes in the elastic range $0\leqslant \gamma_0
\leqslant 0.065$. Note that the data for the undeformed glass
($\gamma_0=0$) are the same in all three panels in
Fig.\,\ref{fig:poten_amp_3pan}. It can be seen that at the low
temperature $T_{LJ}=0.01\,\varepsilon/k_B$, the potential energy
level for the poorly annealed glass in the absence of deformation
remains nearly unchanged during the time interval $1400\,T$. As
indicated in Fig.\,\ref{fig:poten_amp_3pan}, the deformation
procedure includes either periodic shear along one plane or
alternating shear along two and three directions.   As expected, for
each deformation protocol, the system reaches lower energy states
when samples are deformed with larger strain amplitudes. Most
importantly, it can be observed in Fig.\,\ref{fig:poten_amp_3pan}
that for each value of $\gamma_0$, the potential energy curves
become lower when an additional shear orientation is introduced in
the deformation protocol. We further comment that the energy curves
are nearly the same (within fluctuations) for $\gamma_0=0.06$ and
$0.065$ in the panels (b) and (c) in Fig.\,\ref{fig:poten_amp_3pan},
possibly because the critical strain amplitude for these two
protocols is slightly smaller than in the case of periodic shear
along the $xz$ plane, shown in Fig.\,\ref{fig:poten_amp_3pan}\,(a).


\vskip 0.05in

There are several remarks to be made regarding the influence of
temperature and strain amplitude on the potential energy series and
the yielding transition.   The results of test simulations using
three different deformation protocols at the larger strain amplitude
$\gamma_0=0.07$ have shown that the flow becomes localized typically
within the first 300 shear cycles, leading to higher energy levels
in steady state (not shown).   An accurate estimation of the
critical value of the strain amplitude for different deformation
protocols is not the main focus of the present study and it might be
addressed in the future.   An example of the potential energy
dependence on time near the critical strain amplitude during cyclic
shear along a single plane at $T_{LJ}=0.1\,\varepsilon/k_B$ was
reported in Fig.\,2 of the previous study~\cite{Priezjev18a}. We
also comment that the yielding transition was not detected during
the first 600 shear cycles with the strain amplitude $\gamma_0=0.07$
at the temperature $T_{LJ}=0.01\,\varepsilon/k_B$~\cite{Priezjev18}.
In that study, however, the preparation procedure and the
temperature control were different; namely, the dissipative particle
dynamics thermostat versus the Nos\'{e}-Hoover thermostat, which
might lead to delay in the onset of the yielding transition.

\vskip 0.05in

\begin{figure}[t]
\includegraphics[width=12.0cm,angle=0]{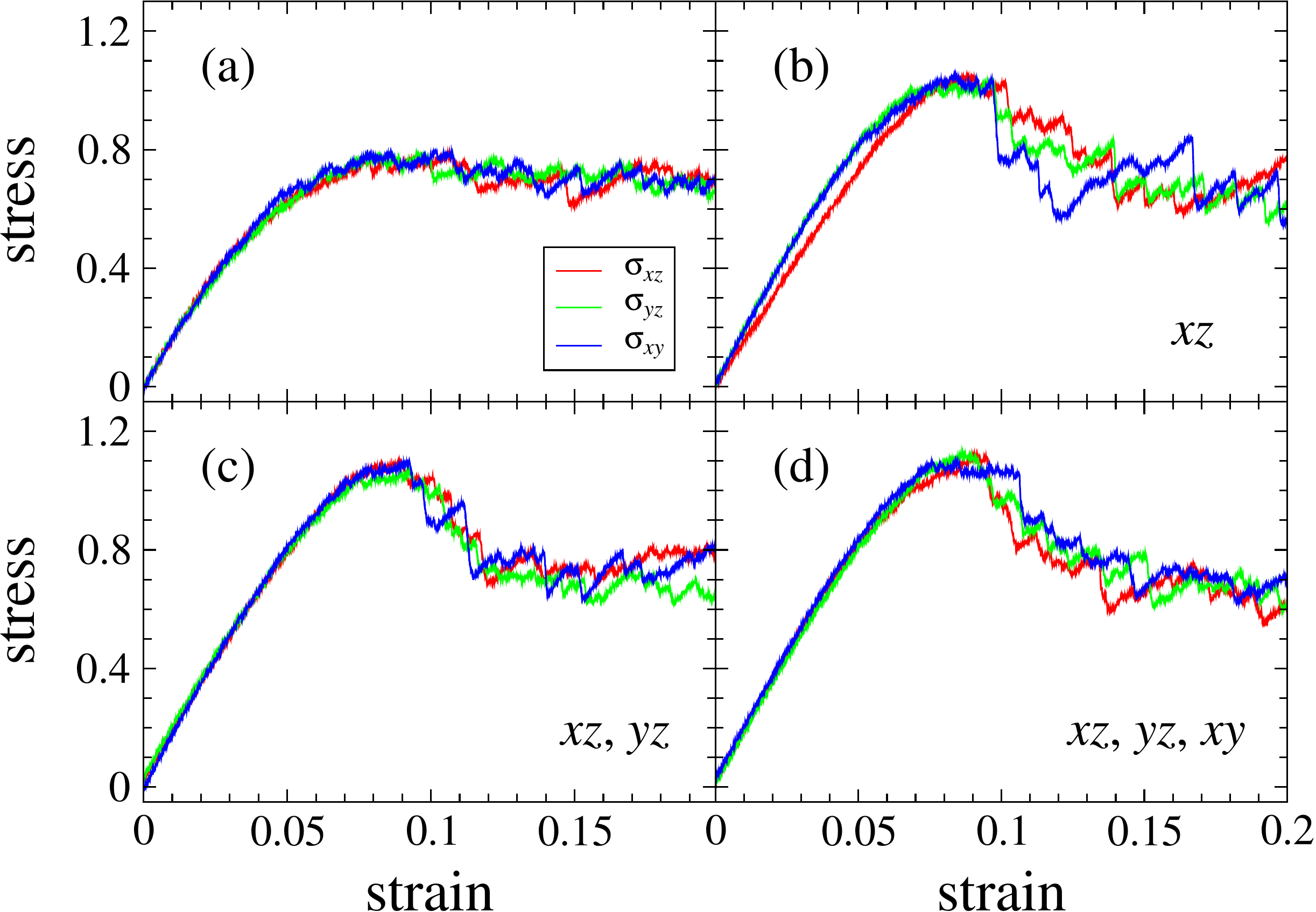}
\caption{(Color online) The shear stress (in units of
$\varepsilon\sigma^{-3}$) as a function of strain during steady
shear with the rate $\dot{\gamma}=10^{-5}\,\tau^{-1}$.  The plane of
shear is parallel to the $xz$ plane (red), the $yz$ plane (green),
and the $xy$ plane (blue). The samples were (a) at mechanical
equilibrium during $1400\,T$, (b) periodically sheared along the
$xz$ plane, (c) sheared along the $xz$ and $yz$ planes, and (d)
deformed along the $xz$, $yz$, and $xy$ planes. See text for
details. }
\label{fig:stress_strain}
\end{figure}

The effect of cyclic annealing on mechanical properties was
investigated by imposing steady shear with the rate of
$10^{-5}\,\tau^{-1}$ at constant volume and temperature
$T_{LJ}=0.01\,\varepsilon/k_B$. Fig.\,\ref{fig:stress_strain} shows
the stress-strain curves after aging and cyclic loading with the
strain amplitude $\gamma_0=0.06$. In each case, after $1400\,T$, the
samples were relaxed at mechanical equilibrium during $10^4\tau$ and
then steadily strained along the $xz$, $yz$, and $xy$ planes in
order to examine the stress response and mechanical anisotropy. It
can be seen in Fig.\,\ref{fig:stress_strain}\,(a) that in the
absence of cyclic loading, the shear stress remains isotropic and it
monotonically increases to a plateau level, as expected for an
amorphous material prepared with a relatively fast cooling rate.  As
is evident from Fig.\,\ref{fig:stress_strain}\,(b), after cyclic
loading along the $xz$ plane, the stress response shows a pronounced
yielding peak, and the shear stress $\sigma_{xz}$ becomes smaller
than $\sigma_{yz}$ and $\sigma_{xy}$.   In other words, the cyclic
loading along a certain plane leads to a reduced stress along that
plane compared to the other directions, while the height of the
stress overshoot remains insensitive to the shear direction.

\vskip 0.05in

\begin{figure}[t]
\includegraphics[width=12.0cm,angle=0]{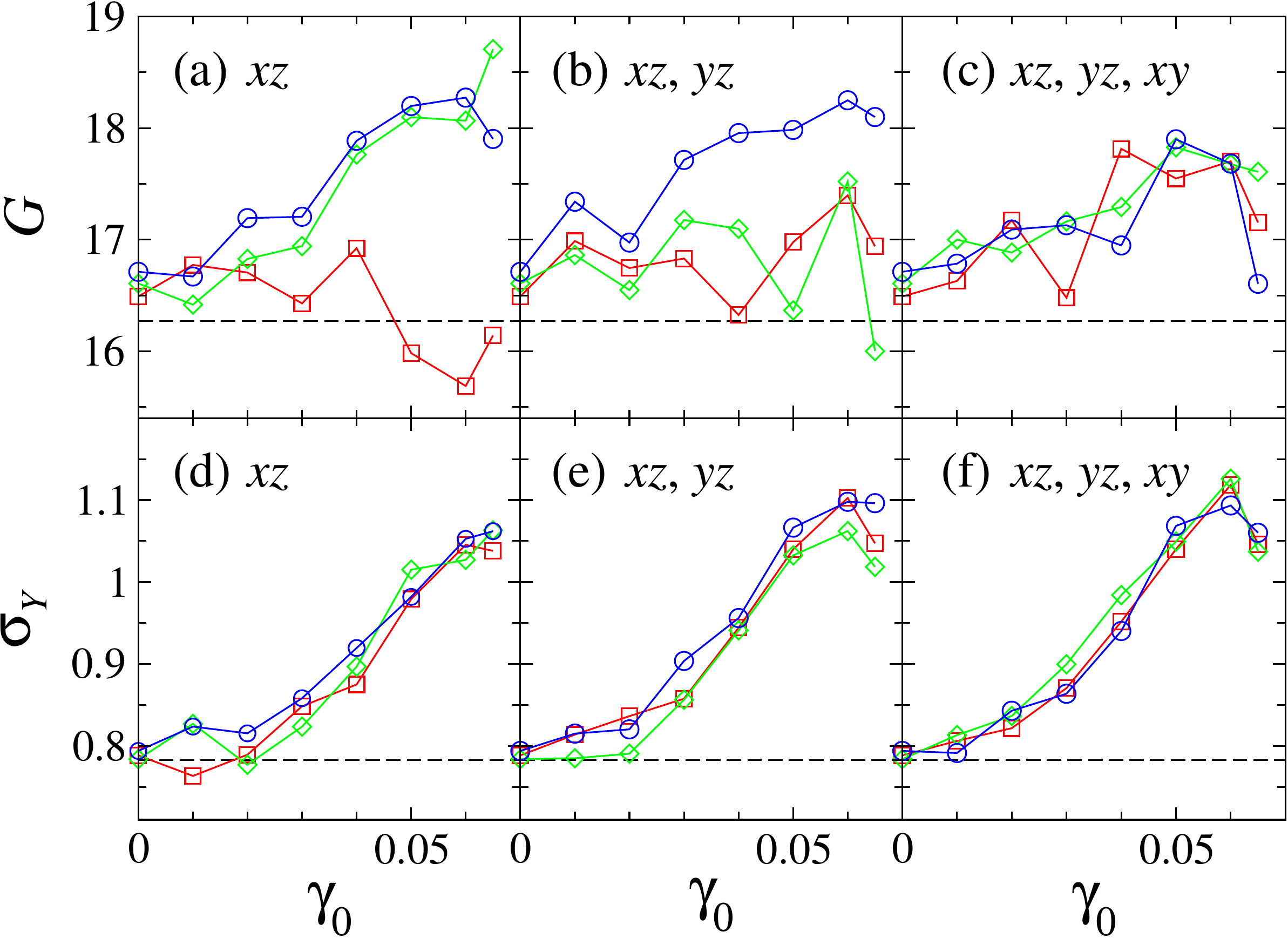}
\caption{(Color online) The shear modulus $G$ (in units of
$\varepsilon\sigma^{-3}$, upper panels) and the yielding peak
$\sigma_Y$ (in units of $\varepsilon\sigma^{-3}$, lower panels)
measured after 1400 shear cycles with the strain amplitude
$\gamma_0$. The annealing protocols include: (a, d) periodic shear
along the $xz$ plane, (b, e) alternating shear along the $xz$ and
$yz$ planes, and (c, f) alternating shear along the $xz$, $yz$, and
$xy$ planes. Both $G$ and $\sigma_Y$ are computed from the
stress-strain curves during steady shear along the $xz$ plane
($\square$), along the $yz$ plane ($\lozenge$), and along the $xy$
plane ($\circ$). The horizontal dashed lines denote the data before
periodic loading. }
\label{fig:Y_G_vs_gam0}
\end{figure}

The stress-strain curves for the samples that were cyclically
deformed with alternating shear along two and three directions are
presented in panels (c) and (d) of Figure\,\ref{fig:stress_strain}.
In both cases, the shape of the curves remains very similar,
although one can notice a slight increase in the height of the
yielding peak when the deformation protocol consists of alternating
shear along three directions.   This is consistent with the lowest
energy minimum after 1400 cycles at $\gamma_0=0.06$ reported in
Fig.\,\ref{fig:poten_amp_3pan}\,(c).   It can also be observed in
Fig.\,\ref{fig:stress_strain}\,(c) that the slope of the
stress-strain dependence is larger along the $xy$ plane, which was
not deformed during cyclic loading.   Furthermore, the variation of
the shear modulus and yielding peak as a function of the strain
amplitude is plotted in Fig.\,\ref{fig:Y_G_vs_gam0}.  The data are
somewhat scattered but the general conclusions can be formulated as
follows. First, the height of the yielding peak increases when an
additional shear orientation is introduced in the cyclic loading
protocol. Second, the shear modulus is larger along the shear
directions that were not used during the cyclic deformation.

\vskip 0.05in


We next discuss the analysis of nonaffine displacements during shear
deformation and provide a visual comparison of strained samples that
were prepared with different cyclic loading protocols. By
definition, the nonaffine displacement of an atom $i$ with respect
to its neighbors can be computed using the matrix $\mathbf{J}_i$,
which linearly transforms the group of neighboring atoms during the
time interval $\Delta t$ and minimizes the following expression:
\begin{equation}
D^2(t, \Delta t)=\frac{1}{N_i}\sum_{j=1}^{N_i}\Big\{
\mathbf{r}_{j}(t+\Delta t)-\mathbf{r}_{i}(t+\Delta t)-\mathbf{J}_i
\big[ \mathbf{r}_{j}(t) - \mathbf{r}_{i}(t)    \big] \Big\}^2,
\label{Eq:D2min}
\end{equation}
where the sum is over the nearest-neighbors within the distance of
$1.5\,\sigma$ from the location of the $i$-th atom at
$\mathbf{r}_{i}(t)$.   It was originally shown that the local shear
transformations in disordered solids can be accurately identified
using a sufficiently large threshold value of the nonaffine
measure~\cite{Falk98}.  More recently, it was found that during
cyclic deformation of amorphous alloys, the typical size of clusters
of atoms with large nonaffine displacements becomes progressively
smaller when the strain amplitude is below the yielding
point~\cite{Priezjev18,Priezjev18a,NVP18strload}.  By contrast, the
yielding transition in both poorly and well annealed glasses at a
finite temperature is associated with the formation of a
system-spanning shear band after a number of transient
cycles~\cite{Priezjev17,Priezjev18a}.

\vskip 0.05in

\begin{figure}[t]
\includegraphics[width=12.cm,angle=0]{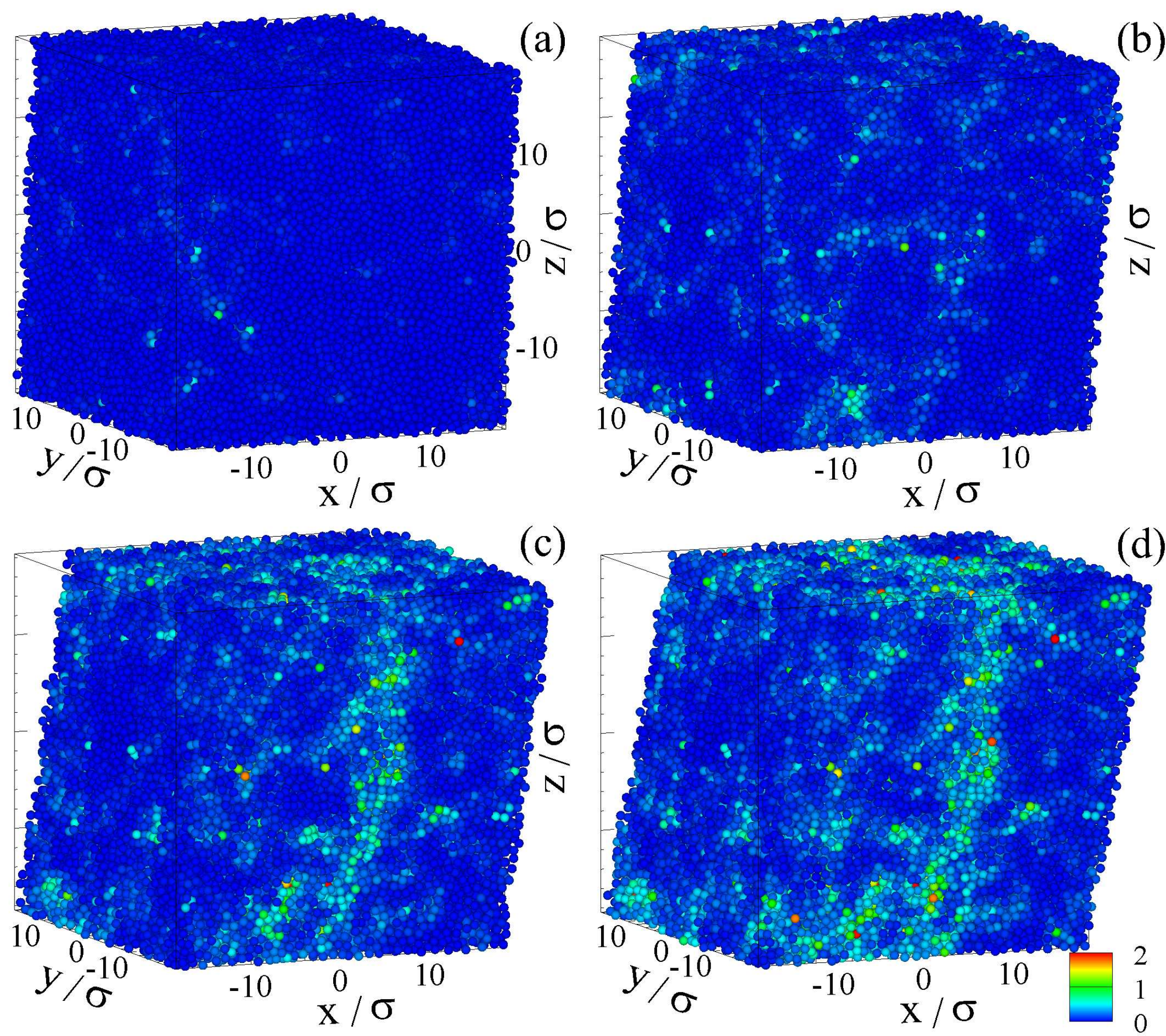}
\caption{(Color online) The sequence of snapshots for the glass aged
during $1400\,T$ at $T_{LJ}=0.01\,\varepsilon/k_B$ and then strained
along the $xz$ plane. The strain rate is $10^{-5}\,\tau^{-1}$ and
the shear strain is (a) $0.05$, (b) $0.10$, (c) $0.15$, and (d)
$0.20$. The color indicates the value of the nonaffine measure with
respect to $\gamma_{xz}=0$, according to the legend. }
\label{fig:snap_amp00_Gxz}
\end{figure}
\begin{figure}[t]
\includegraphics[width=12.cm,angle=0]{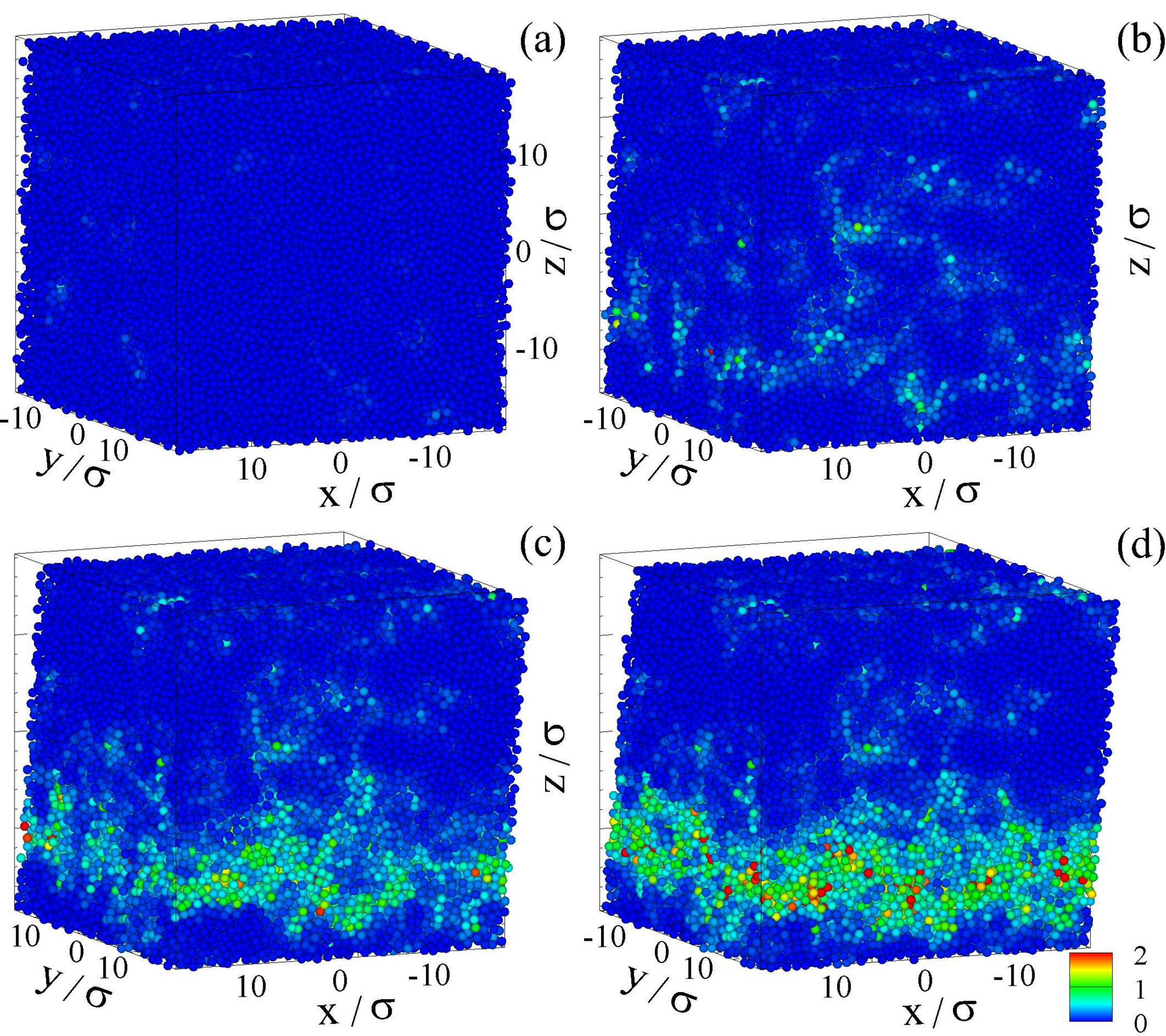}
\caption{(Color online) Instantaneous atomic configurations at the
shear strain (a) $\gamma_{yz}=0.05$, (b) $\gamma_{yz}=0.10$, (c)
$\gamma_{yz}=0.15$, and (d) $\gamma_{yz}=0.20$. The sample was
initially deformed during $1400$ shear cycles along the $xz$ plane.
The color denotes $D^2$ with respect to zero strain. }
\label{fig:snap_xz_amp06_Gyz}
\end{figure}
\begin{figure}[t]
\includegraphics[width=12.cm,angle=0]{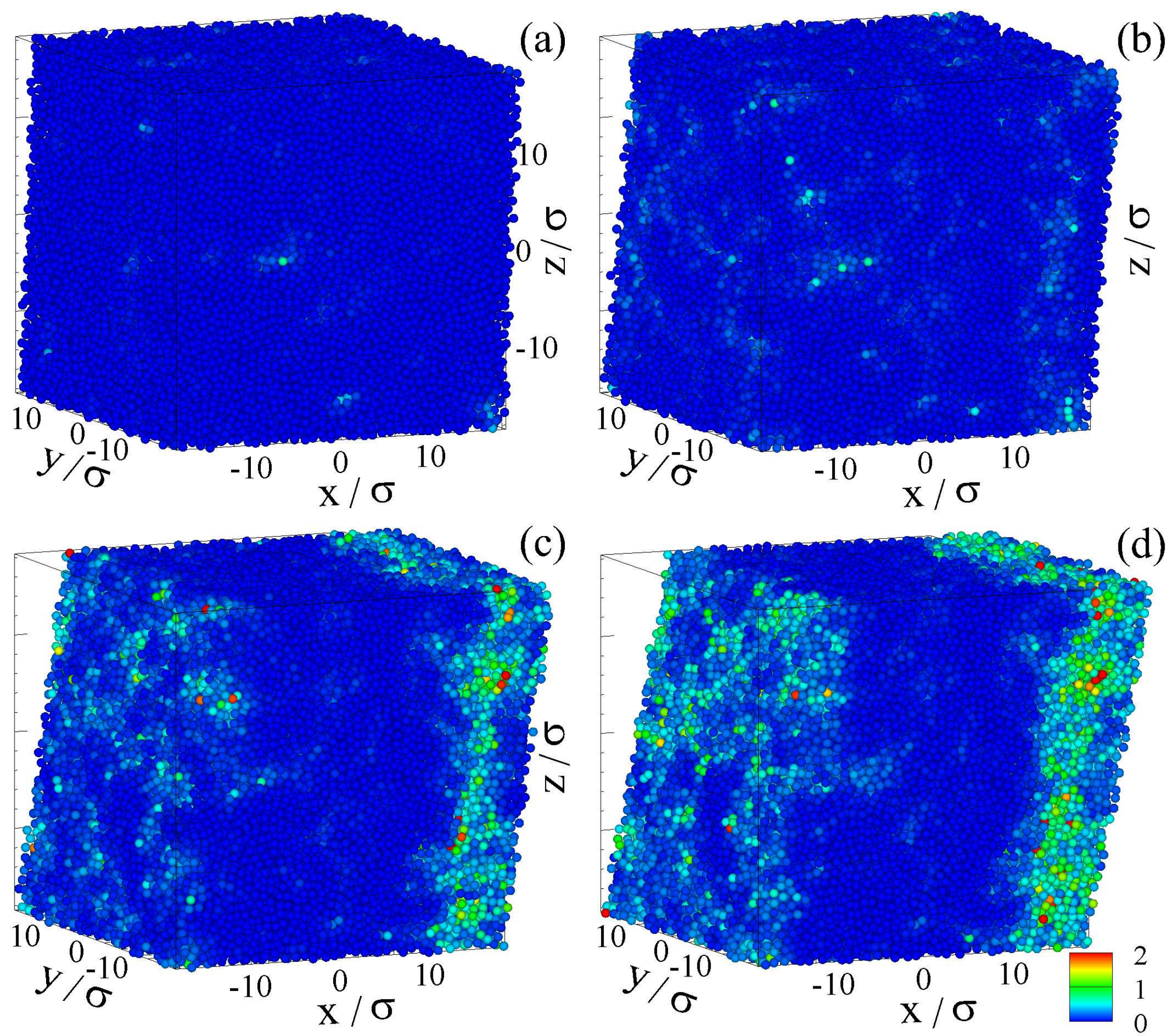}
\caption{(Color online) A series of snapshots during steady shear
along the $xz$ plane with the rate of $10^{-5}\,\tau^{-1}$. The
shear strain is (a) $\gamma_{xz}=0.05$, (b) $0.10$, (c) $0.15$, and
(d) $0.20$. The shear strain was induced after 1400 alternating
shear cycles along the $xz$ and $yz$ planes. The nonaffine measure
$D^2$ is indicated according to the legend.}
\label{fig:snap_xz yz_amp06_Gxz}
\end{figure}
\begin{figure}[t]
\includegraphics[width=12.cm,angle=0]{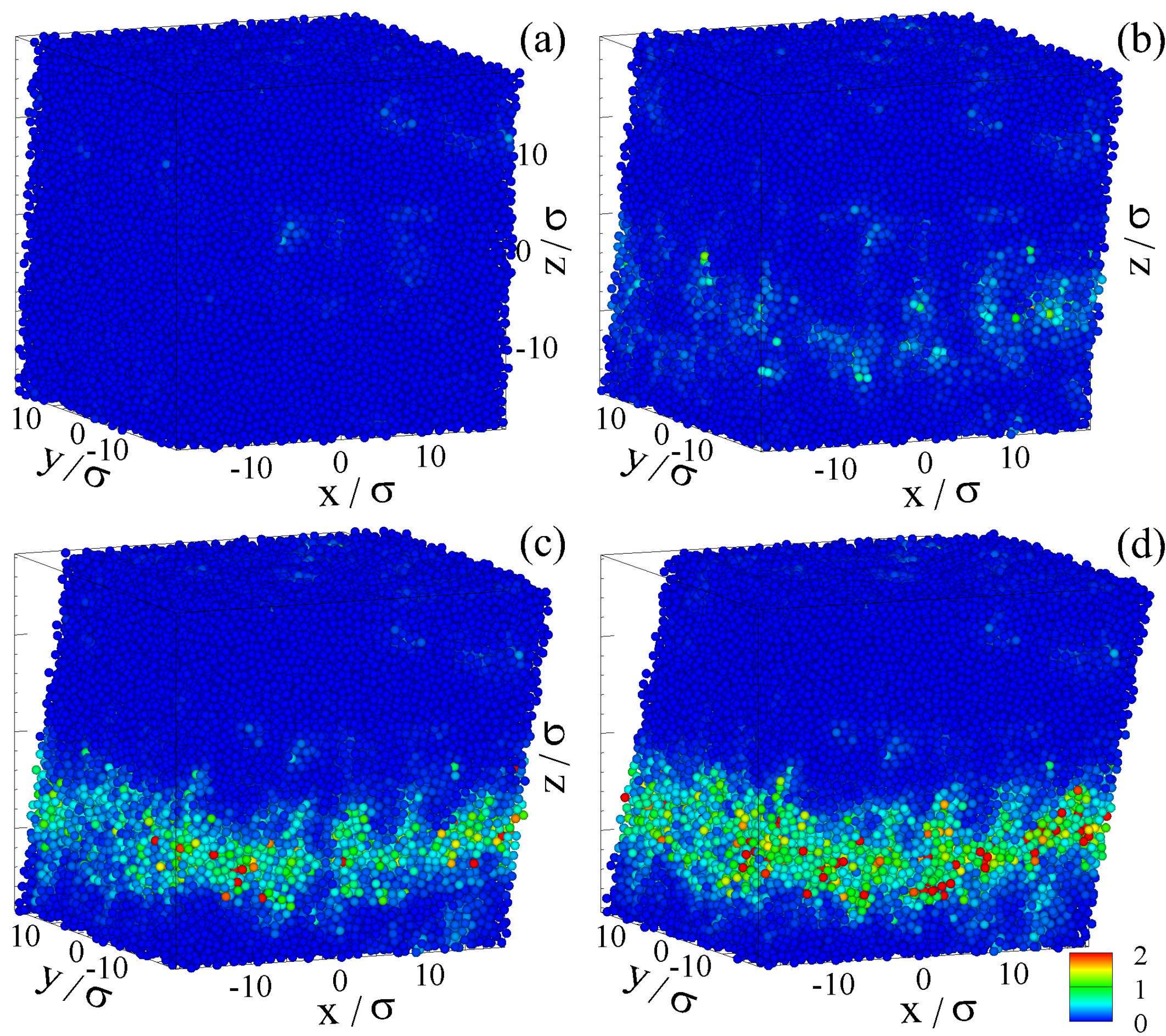}
\caption{(Color online) Four snapshots of the strained glass along
the $xz$ plane with the rate $10^{-5}\,\tau^{-1}$. The strain is (a)
$\gamma_{xz}=0.05$, (b) $0.10$, (c) $0.15$, and (d) $0.20$. The
annealing protocol consists of alternating shear along the $xz$,
$yz$, and $xy$ planes. The color code for $D^2$ is indicated in the
legend box.}
\label{fig:snap_xyz_amp06_Gxz}
\end{figure}

The representative examples of the spatial evolution of nonaffine
displacements are illustrated in
Figs.\,\ref{fig:snap_amp00_Gxz}-\ref{fig:snap_xyz_amp06_Gxz} for
selected stress-strain curves shown in
Fig.\,\ref{fig:stress_strain}.   In all cases, the nonaffine measure
is computed with respect to zero strain and indicated by the color
according to the legend. As shown in Fig.\,\ref{fig:snap_amp00_Gxz},
the plastic deformation of the aged sample is initially rather
homogeneous, and the flow becomes localized only when
$\gamma_{xz}=0.20$.  In contrast, after cyclic loading along the
$xz$ plane, the nucleation of the shear band is apparent right after
the yielding peak at $\gamma_{yz}=0.10$ and it becomes wider with
increasing strain, as shown in Fig.\,\ref{fig:snap_xz_amp06_Gyz}.
Note that in this case the glass is strained along the $yz$ plane,
which is perpendicular to the plane of cyclic shear.  Similar
patterns are evident in strained samples that were cyclically loaded
with alternating shear along two (see Fig.\,\ref{fig:snap_xz
yz_amp06_Gxz}) and three (see Fig.\,\ref{fig:snap_xyz_amp06_Gxz})
directions.  Altogether, these results indicate that cyclic loading
within the elastic regime leads to the increase in strength and
pronounced shear localization.

\section{Conclusions}

In summary, a comparison of different cyclic deformation protocols
aiming to access more relaxed states in disordered solids was
performed using molecular dynamics simulations. We considered a
model glass represented by a strongly non-additive binary mixture,
which was first rapidly cooled from the liquid state to a very low
temperature and then subjected to cyclic loading within the elastic
regime. The deformation protocols include cyclic shear either along
a single plane or periodically alternated along two or three
mutually perpendicular planes. Our main conclusion is that an
additional orientation of the shear plane during the oscillatory
shear deformation results in more relaxed states, provided that the
strain amplitude is smaller than the yield strain. Thus, the lowest
potential energy level is attained in the case of alternating
orientation of the shear plane along all three spatial dimensions.
After cyclic loading, the mechanical properties were probed in
relaxed samples steadily strained along different directions.
Interestingly, the shear modulus tends to be larger along the shear
planes that were not used in the cyclic deformation protocol.  The
analysis of the stress response indicated that the peak height of
the stress overshoot increases in glasses cycled with larger strain
amplitudes.

\section*{Acknowledgments}

Financial support from the National Science Foundation (CNS-1531923)
is gratefully acknowledged. The author would like to thank C.
Reichhardt for drawing attention to the problem of cyclic loading
with alternating shear orientation. The article was prepared within
the framework of the HSE University Basic Research Program and
funded in part by the Russian Academic Excellence Project `5-100'.
The numerical simulations were performed at Wright State
University's Computing Facility and the Ohio Supercomputer Center.
The molecular dynamics simulations were carried out using the LAMMPS
software developed at Sandia National Laboratories~\cite{Lammps}.


%

%

%

%
%

%
%

%
%

%

%

%

\bibliographystyle{prsty}

\end{document}